\documentclass[12pt,fleqn]{article} 
\usepackage{graphicx} 
\usepackage{amsfonts} 
\usepackage{amssymb} 
\usepackage{epsfig}
\usepackage{cite}
\usepackage{user5}  

\newcommand{\rsm}{\ensuremath{{\rm SM}}}
\newcommand{\rmssm}{\ensuremath{{\rm MSSM}}}
\newcommand{\rsusy}{\ensuremath{{\rm SUSY}}}

\newcommand{\sa}{\ensuremath{\sin\alpha}}
\newcommand{\ca}{\ensuremath{\cos\alpha}}

\newcommand{\sbt}{\ensuremath{\sin\beta}}
\newcommand{\cbt}{\ensuremath{\cos\beta}}
\newcommand{\tbeta}{\ensuremath{\tan\beta}}
\newcommand{\ma}{\ensuremath{M_{A^0}}}
\newcommand{\MH}{\ensuremath{M_{H^0}}}
\newcommand{\Mh}{\ensuremath{M_{h^0}}}
\newcommand{\mz}{\ensuremath{M_{Z}}}
\newcommand{\mw}{\ensuremath{M_{W}}}

\newcommand{\msto}{m_{\tilde{t}_1}}
\newcommand{\mstt}{m_{\tilde{t}_2}}

\def\cw{c_{{\scriptscriptstyle W}}}
\newcommand{\mhtree}{M^{2 \,{\mathrm{tree}}}_{h^0}}

\newcommand{\be}{\begin{equation}}
\newcommand{\ee}{\end{equation}}
\newcommand{\bea}{\begin{eqnarray}}
\newcommand{\eea}{\end{eqnarray}}

\newcommand{\VL}{\left( \begin{array}{c}}
\newcommand{\VR}{\end{array} \right)}
\newcommand{\ML}{\left( \begin{array}{cc}}
\newcommand{\MLd}{\left( \begin{array}{ccc}}
\newcommand{\MLv}{\left( \begin{array}{cccc}}
\newcommand{\MR}{\end{array} \right)}

\begin{document} 
\begin{titlepage} 
\hfill{} 
\begin{tabular}{l} 
KA-TP-20-2001 \\ 
hep-ph/0108245 \\ 
\end{tabular} 
\vspace*{1cm}\\
\begin{center}
\textbf{\large Yukawa coupling quantum corrections to the self 
couplings \\[0.2cm]
of the lightest MSSM Higgs boson}\vspace*{1cm}\\
{\par\centering 
Wolfgang Hollik and Siannah Pe{\~n}aranda\vspace*{0.3cm}~\footnote{electronic addresses:
hollik@particle.uni-karlsruhe.de, siannah@particle.uni-karlsruhe.de}\\
\par} 
{\par\centering 
\textit{Institut f\"{u}r Theoretische Physik, Universit\"{a}t Karlsruhe}\\ 
\textit{Kaiserstra\ss{}e 12, D--76128 Karlsruhe, Germany }}
\end{center}
\vspace*{2cm}
{\par\centering\textbf{\large Abstract}\\ 
\vspace*{0.4cm}
\par} 
\noindent A detailed analysis of the top-quark/squark quantum corrections to
the lightest CP-even Higgs boson $(h^0)$ self-couplings is presented
in the MSSM. By
considering the leading one-loop Yukawa-coupling contributions of 
${\cal O} (m_t^4)$, we discuss the decoupling
behaviour of these corrections when the top-squarks are heavy compared
to the electroweak scale. As shown analytically and numerically,
the large corrections  can almost completely be absorbed into the 
$h^0$-boson mass.
Our conclusion is that the $h^0$ self-couplings remain similar to the coupling
of the $\rsm$ Higgs boson for a heavy top-squark sector.
\end{titlepage} 

\vfill
\clearpage
%%%%%%%%%%%%%%%%%%%%%%%%%%%%%%%%%%%%%%%%%%%%%%%%%%%%%%%%%%%%%%%%%%%%%%%%

\renewcommand{\thefootnote}{\arabic{footnote}}
\setcounter{footnote}{0}

%%%%%%%%%%%%%%%%%%%%%%%%%%%%%%%%%%%%%%%%%%%%%%%%%%%%%%%%%%%%%%%%%%%%%%%%%%%%
\pagestyle{plain}

\section{Introduction} 

The Standard Model ($\rsm$) with minimal Higgs-field content 
could turn out not to be basic
theoretical framework for describing electroweak symmetry brea\-king.
In recent years supersymmetry (\rsusy) has become one of the most promising
theoreti\-cal ideas beyond the SM.
The Minimal Supersymmetric Standard Model
($\rmssm$)~\cite{MSSMHiggssector} is the simplest supersymme\-tric
extension of the $\rsm$ and at least as successful as the SM
to describe the experimental data~\cite{jellis}.

The Higgs sector of the MSSM~\cite{BibliaHiggs}
involves two scalar doublets, $H_1$ and $H_2$, 
in order to 
give masses to up- and down-type fermions in a way consistent with 
supersymmetry. After spontaneous symmetry breaking, 
induced through the neutral
components of $H_1$ and $H_2$ with vacuum expectation values
$v_1$ and $v_2$, respectively, 
the $\rmssm$ Higgs sector contains five 
physical states: two neutral CP-even scalars
($h^0$ and $H^0$), one CP-odd pseudoscalar ($A^0$), and two charged-Higgs 
states ($H^{\pm}$). 
The Higgs potential of the $\rmssm$ is constrained by 
SUSY~\cite{BibliaHiggs}: 
all quartic 
coupling constants are related to the electroweak gauge coupling
constants, thus
imposing various restrictions on the tree-level Higgs-boson masses, 
couplings and mixing angles. 
In particular, all tree-level Higgs parameters can be determined 
in terms of the mass of the CP-odd Higgs boson, $\ma$, 
and the ratio 
$\tbeta = {v_2}/{v_1}$. The other masses and 
the mixing angle $\alpha$ are then fixed, and the trilinear and
quartic self-couplings of the physical Higgs particles can be predicted. 
The knowledge of the Higgs-boson self-couplings will be essential 
for establishing the Higgs potential and thus the
Higgs mechanism as the basic mechanism for generating the masses of the
fundamental particles~\cite{Reports}.

The tree-level relations among the Higgs-boson masses
in the MSSM acquire relevant 
radiative corrections, dominated by top-quark/squark
loops~\cite{RadCorr1}. 
Extensive effort has been devoted to progressive
refinements of the radiative corrections for the Higgs-boson 
masses, with special emphasis on the prediction of the
lightest MSSM Higgs-boson mass $\Mh$, 
by using different techniques:
renormalization group equations~\cite{RGEHiggs,Carena:1996wu}, diagrammatic 
computations~\cite{Higgsoneloop,Dabelstein,Heinemeyer,Hollikmass},
and a combination of both~\cite{CarenaHollik}.
Besides the mass spectrum, the loop corrections influence the production
cross sections and decay branching ratios~\cite{HHRW}.

Moreover, large quark/squark-loop
corrections affect also  the self-couplings of the neutral Higgs 
particles~\cite{RadCorCouplings,osland,Djouadi,Djouadi2,OtrosH}. 
The loop contributions modify the mixing angle $\alpha$ for
the neutral CP-even mass eigenstates and alter the 
triple and quartic Higgs-boson self-couplings. 
Since in the limit of a heavy $A^0$ boson the light-Higgs 
couplings to gauge bosons and fermions become very close to those
of the SM, 
quantum effects can play a crucial role to distinguish between
a SM and a MSSM light Higgs boson. In this context, also the 
investigation of the decoupling behaviour of quantum effects in the
Higgs self-interaction are of interest.

In this paper we are concerned with the one-loop corrections 
to the self-couplings of the lightest CP-even MSSM Higgs boson, $h^0$.
As a first step, we  
analyze here the leading one-loop Yukawa contributions
of~${\cal O}(m_t^4)$ to the $h^0$
one-particle irreducible (1PI)
Green functions, which yield, besides the Higgs-boson 
mass corrections, the effective triple and quartic self-couplings.  
We study, both numerically and analytically, 
the asymptotic behaviour of
these corrections in the limit of heavy top squarks, with masses
large as compared to the electroweak scale, and 
discuss the decoupling behaviour of a heavy top-squark system
in the Higgs sector, which becomes particularly interesting 
for large values of $M_A$ when $h^0$ is the only light Higgs particle.
The corresponding analysis of all the one-loop 
contributions from the Higgs sector 
to the $h^0$ self-couplings will be presented elsewhere~\cite{Prepara2}.

The decoupling properties of the 
one-loop radiative corrections to various observables have been extensively
studied in the literature
\cite{Drees:1990dx,Sola1,eff,Haberetal,CPR,Sola2,MJHtb,Djouadi:1998pb,HollikYO,Distin}.
Concerning Higgs physics, 
it is well known that the SUSY one-loop corrections 
to the couplings of Higgs bosons to $b$-quarks can be 
significant for large values of $\tan \beta$,
and that they do not decouple, in general, in
the limit of a heavy supersymmetric spectrum~\cite{Sola1,eff,Haberetal,MJHtb}.
Conversely, it has been shown that all the non-standard particles in the
$\rmssm$ decouple from low-energy electroweak gauge-boson 
physics~\cite{TesisS,Tesis-Proc}.

This paper is organized as follows: In section~\ref{sec:Spectrum} 
notations are given and a brief collection of formulae for the 
top-squark sector and for the Higgs sector in the 
$\rsm$ and $\rmssm$, describing the asymptotic limits
being considered here. The asymptotic results for the 
top-quark/squark contributions to the vertex functions of the $h_0$ 
and a discussion of decoupling properties are contained
in section~\ref{sec:selfcouplings}. 
A more explicit discussion of
the ${\cal O}(m_t^4)$ radiative corrections to the
trilinear and quartic $h^0$-boson self-couplings is given in
section~\ref{sec:tri}, with a short summary in
section~\ref{sec:conclu}.

\section{Particle spectrum and decoupling limit}
\label{sec:Spectrum}

\subsection{MSSM squark sector} 
\label{sec:squarksector}

Since we are dealing with the leading quark/squark contributions to the
Higgs sector, we briefly describe the input from the top-squark sector 
and specify the asymptotic limits for the subsequent discussion.
For simplicity we assume  that there is no
intergenerational flavour mixing. 
The tree-level $\tilde{t}$ squared-mass matrix reads
\begin{equation}
{\cal M}_{\tilde{t}}^2 =\left(\begin{array}{cc}
M_L^2 &
m_t\, X_t\\ m_t\, X_t & M_R^2
\end{array} \right)\,,
\label{eq:stopmatrix}
\end{equation}
where 
\bea 
M_L^2&=&M_{\tilde Q}^2+m_t^2 + \left(
       \frac{1}{2}- \frac{2}{3} s_W^2 \right)\,\mz^2\,\cos{2\beta}\,, 
       \nonumber \\
M_R^2&=&M_{\tilde U}^2+m_t^2+
       \frac{2}{3} \,s_W^2\,\mz^2\, \cos{2\beta} \, , 
        \nonumber \\
X_t&=&A_t-\mu\cot\beta\,,
\label{eq:MLRtb}
\eea
and $s_W\equiv \sin\theta_W$. The parameters
$M_{\tilde Q}$ and $M_{\tilde U}$ are the soft-SUSY-breaking masses,
$A_{t}$ is the corresponding soft-SUSY-breaking trilinear
coupling, and $\mu$ is the bilinear coupling of the two Higgs doublets. 

Diagonalizing the $\tilde t$-mass matrix (\ref{eq:stopmatrix}) yields the mass
eigenvalues  $m^2_{\tilde t_{1,2}}$ and the $\tilde t$-mixing angle 
$\theta_{\tilde t}$, relating the current eigenstates to the mass
eigenstates,
\begin{equation}
\left( \begin{array}{c}
\tilde t_1\\ \tilde t_2 
\end{array}\right) = 
\left( \begin{array}{cc}
\cos \theta_{\tilde t} & - \sin \theta_{\tilde t} \\  \sin
\theta_{\tilde t} & \cos \theta_{\tilde t}
\end{array}\right)
\left( \begin{array}{c}
\tilde t_L\\ \tilde t_R 
\end{array}\right)\,.
\label{eq:squarkrotation}
\end{equation}
The corresponding stop-mass eigenvalues, with the convention 
$m_{\tilde t_1}> m_{\tilde t_2}$, are given by
\bea
    m^2_{\tilde t_{1,2}} = \frac{1}{2}\left[ M_L^2 + M_R^2
    \pm \sqrt{ (M_L^2 - M_R^2)^2 + 4 m_t^2 X_t^2 } \right]\,,
\eea
and the mixing angle $\theta_{\tilde t}$ is determined via
\be \label{thetab}
    \cos 2 \theta_{\tilde t} = \frac{M_L^2 - M_R^2}
    {m^2_{\tilde t_1}-m^2_{\tilde t_2}}\,\,,\,\,\,\,\,\,\,
    \sin 2 \theta_{\tilde t} =
    \frac{2 m_t X_t}{m^2_{\tilde t_1}-m^2_{\tilde t_2}}\,.
\ee

With respect to our analysis of decoupling, we consider  
the asymptotic limit in which the $\tilde{t}$ masses
 are very large as compared to the external momenta and to 
the electroweak scale,
\be
\label{eq:limitgeneral}
\msto^2\,,\mstt^2  \gg \mz^2\,,\Mh^2\,.
\ee 
Since, however, the asymptotic behaviour of one-loop 
integrals with internal $\tilde{t}$ lines
depend on the relative size of the top-squark
masses in the loop propagators, more specific assumptions have to be made.
The only two internal masses that can be different in the loop diagrams
are $\msto\,,\mstt$ (see Fig.~\ref{fig:fdiagrams}
for the generic diagrams considered here). For the discussion 
in section~\ref{sec:selfcouplings} we assume 
that these two  $\tilde{t}$  masses are heavy but close to each other, i.e. 
\be
\label{eq:limit}
|\msto^2-\mstt^2| \ll |\msto^2+\mstt^2|\,.
\ee
A detailed discussion of this limit can be found
in~\cite{TesisS}. Another possible scenario is the case
where the stop mass splitting is of the order of 
the $\rsusy$ mass scale, $M_{\tilde Q}$,
\be
\label{eq:limit2}
 |\msto^2-\mstt^2| \simeq |\msto^2+\mstt^2|\,\,,
\ee
which will be considered in section~\ref{sec:tri}.

\subsection{SM and MSSM Higgs sector}
\label{sec:selH}

The electroweak gauge bosons and the fundamental matter particles 
of the $\rsm$ acquire
their masses through the interaction with the Higgs field.
To establish the Higgs mecha\-nism experimentally, the
characteristic self-interaction potential of the $\rsm$,
$V = \lambda \left(|\varphi|^2 -\textstyle{\frac{1}{2}} v^2 \right)^2$, 
with a minimum at $\langle \varphi \rangle_0 = v/\sqrt{2}$, must be
reconstructed once the Higgs particle will be discovered. This
task requires the measurement of the trilinear and
quartic self-couplings of the Higgs boson, $H_{\rsm}$. The self-couplings are
uniquely determined in the $\rsm$ by the mass of the Higgs
boson, which is related to the quartic coupling $\lambda$ by $M_H
= \sqrt{2\lambda} v$. Introducing the physical Higgs field $H=H_{\rsm}$ in the
neutral component of the doublet, $\varphi^0 = (v+H)/\sqrt{2}$, the
trilinear and quartic vertices of the Higgs field $H$ 
can be derived from the potential $V$, yielding
\be
\label{eq:coupSM}
\lambda_{HHH} = \frac{3g M_H^2}{2\,\mw} 
              = \frac{3 M_H^2}{v}\,, \,\,\,\quad
\lambda_{HHHH} = \frac{3g^2 M_H^2}{4\,\mw^2} 
               = \frac{3 M_H^2}{v^2} \,,
\ee
with the ${\rm SU(2)_L}$ gauge coupling $g$.

In the $\rmssm$, 
the  2-doublet Higgs potential is given by~\cite{BibliaHiggs}
\bea
\label{eq:Higgspot}
V &=& m_1^2 H_1\bar{H}_1 + m_2^2 H_2\bar{H}_2 + m_{12}^2\, (\epsilon_{ab}
      H_1^a H_2^b + {\rm {h.c.}})  \nonumber \\
   && \mbox{} + \frac{g'^2 + g^2}{8}\, (H_1\bar{H}_1 - H_2\bar{H}_2)^2
      +\frac{g^2}{2}\, |H_1\bar{H}_2|^2,
\eea
with the doublet fields $H_1$ and $H_2$,
the soft SUSY-breaking terms $m_1, m_2, m_{12}$, and 
the ${\rm SU(2)_L}$ and ${\rm U(1)_Y}$ gauge couplings $g, g'$.

Two parameters, conveniently chosen to be
the CP-odd Higgs-boson mass $\ma$ 
($\ma^2 = m_{12}^2(\tbeta+\cot \beta)$)
and $\tbeta=v_2/v_1$, are sufficient to fix all the other 
parameters of the tree-level Higgs sector.
The two CP-even neutral mass eigenstates are a mixture of the
real neutral $H_1$ and $H_2$ components,
\bea
\VL H^0 \\ h^0 \VR &=& \ML \ca & \sa \\ -\sa & \ca \MR 
\VL  H_2^0 \\ H_1^0 \VR  ,
\label{higgsrotation}
\eea
with the mixing angle $\alpha$ related to $\tbeta$ and $\ma$ by
\be
\tan 2\alpha = \tan 2\beta \, \frac{\ma^2 + \mz^2}{\ma^2 - \mz^2}\,,
\quad - \frac{\pi}{2} < \alpha < 0 \,. 
\ee

The tree-level mass matrix of the neutral CP-even Higgs bosons
can be expressed in terms of $\mz$, $\ma$ and the angle $\beta$ as follows:
\bea
M_{\rm Higgs}^{2, {\rm tree}}
&=& \ML \ma^2 \sin^2 \beta + \mz^2 \cos^2 \beta & -(\ma^2 + \mz^2) \sbt \cbt \\
    -(\ma^2 + \mz^2) \sbt \cbt & \ma^2 \cos^2 \beta + \mz^2 \sin^2 \beta \MR.
\eea

The eigenvalues of $M_{\rm Higgs}^{2, {\rm tree}}$ are the squared 
masses of the two CP-even Higgs scalars, in terms of $\ma$ and $\beta$
given by
\be
M^2_{H^{0}, h^{0}} = \frac{1}{2}
    \left[ \ma^2 + \mz^2 \pm \sqrt{\left(\ma^2 + \mz^2\right)^2 -
           4 M_A^2 \mz^2 \cos^2 2 \beta }\, \right]\, .
\ee

These tree-level predictions for the CP-even
Higgs-boson masses and
mixing angle, however, are subject to large radiative corrections,
with sensitive dependence on the top mass.
Explicit analytical expressions for
the logarithmic and non-logarithmic contributions to $\Mh$,
including the dominant two-loop terms, can
be found in~\cite{Hollikmass}. 

The tree-level trilinear and quartic  $h^0$ couplings 
in the MSSM, which are in the focus of the present work, 
can be written as follows,
\bea
\label{eq:couptree}
\lambda_{hhh}^0 &=&  3 \,\frac{g \mz}{2 \cw}
\cos2\alpha \sin(\beta + \alpha)\,, \nonumber\\
\lambda_{hhhh}^0 &=& 3 \,\frac{g^2}{4 \cw^2}\cos^2 2\alpha\,,
\eea
with $\cw = \cos\theta_W$.

Obviously, they are different from the couplings of the $\rsm$ Higgs boson
(\ref{eq:coupSM}). However, the situation changes 
in the so-called {\it decoupling limit} of the Higgs sector. 
The {\it decoupling limit}, 
studied first in Ref.~\cite{dec}, is, in short, defined 
by considering a large CP-odd Higgs-boson mass $M_A \gg \mz$,
yielding a particular spectrum in the Higgs sector 
with very heavy $H^0$, $H^{\pm}$, $A^0$ bosons obeying
$\ma \simeq \MH \simeq M_{H^{\pm}}$ [up to terms of 
${\cal O}(\frac{\mz}{\ma})$]
and a light $h^0$ boson with a tree-level mass of
$M^{{\mathrm{tree}}}_{h^0}\simeq \mz |\cos2\beta|$. In this limit, 
which also implies $\alpha \rightarrow \beta -\pi/2$,
 one obtains that the self couplings~(\ref{eq:couptree}) tend towards
\be 
\label{treelevelself}
\lambda_{hhh}^0 \simeq
 \,3 \,\frac{g}{2\,\mw} \,\mhtree\, , \quad 
\lambda_{hhhh}^0 \simeq
\,3\, \frac{g^2}{4\,\mw^2} \mhtree \, ,
\ee
and thus the tree-level couplings of the light
CP-even Higgs boson approach the couplings~(\ref{eq:coupSM})  
of a $\rsm$ Higgs boson with the same mass.

Relevant radiative corrections are also expected for the
light CP-even Higgs-boson self-couplings, dominated by 
the top-quark/squark contributions (see the discussions 
in \cite{RadCorCouplings,osland,Djouadi,Djouadi2,OtrosH} 
for trilinear couplings).
In the following we investigate these dominant one-loop
contributions to the $h^0$ self-couplings and analyze 
their behaviour in the {\it decoupling limit}.

\section{Higgs boson self-couplings}
\label{sec:selfcouplings}

\subsection{Leading Yukawa corrections in the asymptotic limit}
\label{sec:yukawa}
 
Here we derive the one-loop leading Yukawa corrections 
from top and stop loop contributions 
to the one-, two-, tree- and four-point vertex
functions of the lightest Higgs boson, $h^0$, 
and study their asymptotic behaviour for a heavy top-squark sector.
The three- and four-point vertex functions correspond to the
$h^0$ self-couplings. The computation has been performed by the
diagrammatic method using {\it FeynArts 3} and {\it FormCalc}~\cite{Hahn}, 
and the results are expressed in terms of the
standard one-loop integrals~\cite{Passarino}.

The general results for the $n$-point vertex functions
can be summarized by the following generic expression,
\begin{equation}
\label{eq:notG} 
\Gamma_{h^{0}}^{\,t,\tilde{t}\,(n)}={\Gamma_{\,0\,\,h^{0}}^{(n)}}
+\Delta \Gamma_{h^{0}}^{\,t,\tilde{t}\,(n)}\,,
\end{equation}
where the subscript $0$ refers to the tree-level functions, 
which correspond directly to the  expressions
for the $h^{0}$ Higgs couplings already given in~(\ref{eq:couptree}). 
The one-loop contributions are summarized 
in $\Delta \Gamma_{h^{0}}^{\,t,\tilde{t}\,(n)}$. In particular, 
$\Delta \Gamma_{h^{0}}^{\,t,\tilde{t}\,(1)}$ is 
the tadpole contribution and  
$\Delta \Gamma_{h^{0}}^{\,t,\tilde{t}\,(2)}$  the 
$h^{0}$ self-energy;
$\Delta \Gamma_{h^{0}}^{t,\tilde{t}\,(3)}$ and
$\Delta \Gamma_{h^{0}}^{t,\tilde{t}\,(4)}$
are the corresponding radiative corrections to the trilinear and quartic 
$h^0$ self-couplings (choosing a
normalization that the Feynman diagrams yield always $-i\, \Gamma$). 

In order to obtain the asymptotic behaviour 
of the one-loop integrals we assume in this section the conditions given
in~(\ref{eq:limitgeneral}),~(\ref{eq:limit}) and use the
asymptotic expressions of the one-loop integrals
presented in~\cite{TesisS}, 
and the appropriate results for the integrals with equal masses in the loop
propagators given in~\cite{IntHaber}.

\begin{figure}[t]
\begin{center}
\epsfig{file=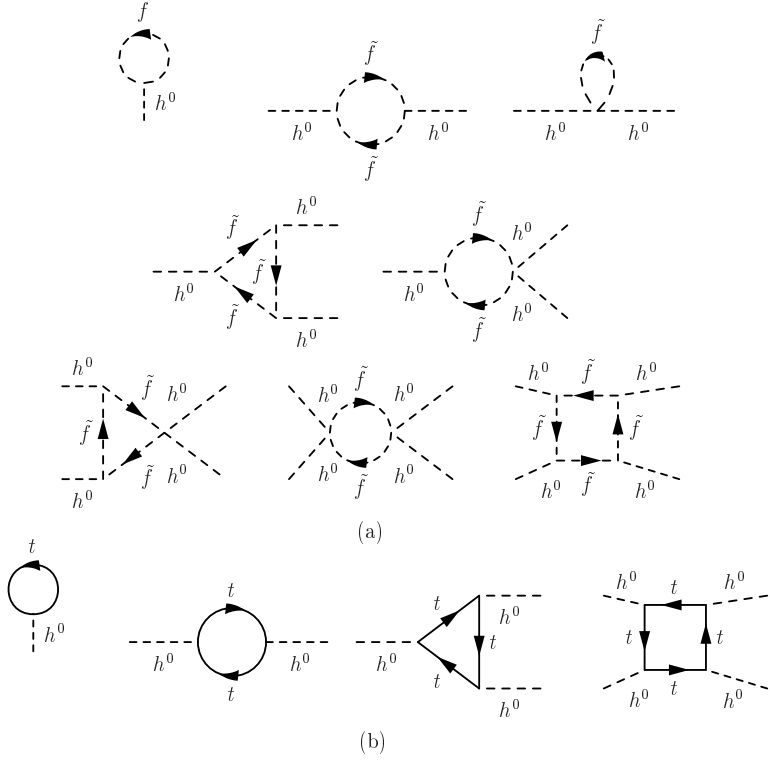,width=12.5cm}
\vspace*{-0.5cm}
\caption{Generic diagrams contributing to the one-, two-, three- and 
four-point 1PI Green functions of the lightest Higgs boson $h^0$ at 
the one-loop level:
\textbf{a)} with top-squark loops, $\tilde f \equiv \tilde{t}_1,
\tilde{t}_2$\,,
\textbf{b)} with top-quark loops.}\label{fig:fdiagrams}
\end{center}
\end{figure}
The diagrams contributing to the $n$-point vertex functions 
$(n=1,...,4)$ from the top-quark/squark sector
are shown generically in Fig.~\ref{fig:fdiagrams}.
The corresponding analytic expressions $\sim {\cal O} (m_t^4)$, 
in the asymptotic limit according 
to~(\ref{eq:limitgeneral})-(\ref{eq:limit}), are given by
\bea
\label{eq:oneloop}
\Delta \Gamma_{h^{0}}^{t,\tilde{t}\,(1)}&=&
\frac{3}{8 \pi^2}\frac{g}{\mw}\,m_t^4\,\left(
\Delta_\epsilon +1-\log\frac{m_t^2}{\mu_{0}^2}\right)\,,\nonumber\\
\Delta \Gamma_{h^{0}}^{t,\tilde{t}\,(2)}&=&
\frac{3}{16 \pi^2}\frac{g^2}{\mw^2}\,m_t^4\,\left(
\Delta_\epsilon +1+\log\frac{\msto^2}{\mu_{0}^2}
+\log\frac{\mstt^2}{\mu_{0}^2}-3\log\frac{m_t^2}{\mu_{0}^2}\right)\,,
\nonumber\\
\Delta \Gamma_{h^{0}}^{t,\tilde{t}\,(3)}&=&
-\frac{3}{16 \pi^2}\frac{g^3}{\mw^3}\,m_t^4\,\left(
2 +3 \log \frac{m_t^2}{\msto \mstt}\right)\,,\nonumber\\
\Delta \Gamma_{h^{0}}^{t,\tilde{t}\,(4)}&=&
-\frac{3}{32 \pi^2}\frac{g^4}{\mw^4}\,m_t^4\,\left(
8 +3 \log \frac{m_t^2}{\msto \mstt}\right)\,,
\eea
where $\mu_{0}$ is the scale of dimensional regularization.
All other terms depending on the external momenta and on
the stop-mass splitting vanish for large values of $m_{\tilde{t}_{1,2}}$.
The three- and four-point functions are UV-finite, whereas
the one- and two-point functions involve a singular 
$\Delta_\epsilon$ term, with
\begin{equation} 
\label{eq:Delta}
\hspace*{0.6cm}  
\displaystyle {\Delta}_\epsilon=\frac{2}{\epsilon }-{\gamma }_{\epsilon}  
+\log (4\pi) \hspace*{0.2cm}, \hspace*{0.2cm} \epsilon = 4-D\,. 
\end{equation} 
All these contributions contain a logarithmic dependence on
the stop masses. These functions are
the only remainder of a heavy stop system 
in the vertex functions of the $h^{0}$ 
and thus summarize all the  potential non-decoupling effects of these
particles in the effective potential for the lightest Higgs boson.
At this point, one could be tempted to conclude that heavy top-squarks 
do not decouple in the Green functions of the 
lightest CP-even Higgs boson of the $\rmssm$ and therefore, in the
$h^{0}$ self-couplings. It is essential, however, to study 
whether those  effects appear in the relations between  
observables~\cite{App-Cara}. 

There are also non-logarithmic finite contributions 
to the three- and four-point functions in~(\ref{eq:oneloop}).
These terms arise from the last two diagrams in 
Fig.~\ref{fig:fdiagrams}. They are also present for the 
Higgs particle ($H_{\rsm}$) in the external legs, instead of $h^{0}$.
Therefore, they do not contribute to the difference between the
$h^{0}$ and $H_{\rsm}$ properties (see next section).

\subsection{Renormalized vertices and decoupling behaviour}
\label{sec:decoupling}

The vertex functions obtained from the set of one-loop diagrams
are in general UV-divergent.
For finite 1PI Green functions and physical observables, 
renormalization has to be performed by adding appropriate counterterms.
For a systematic 1-loop calculation, the free parameters of the Higgs
potential  $m_1^2,\, m_2^2,\, m_{12}^2,\, g,\, g'$
and the two vacua $v_1,\, v_2$ are replaced by renorma\-li\-zed parameters
plus counterterms. This transforms the potential $V$ into $V +\delta V$, 
where V, expressed in terms of the renormalized parameters,
is formally identical to (\ref{eq:Higgspot}), and 
$\delta V$ is the counterterm potential. 
By using the standard renormalization 
procedure~\cite{CPR,Dabelstein} with
$m_i^2 \rightarrow Z_{H_{i}}^{-1} (m_i^2+\delta m_i^2)\,,$
$g \rightarrow Z_{1}^{W} Z_{2}^{W\,-\frac{3}{2}} g\,,$
$g' \rightarrow Z_{1}^{B} Z_{2}^{B\,-\frac{3}{2}} g'\,,$
$v_i \rightarrow Z_{H_{i}}^{1/2}(v_i -\delta v_i)\,,$ and
with field renormalization constants $\delta Z_{H_{i}}$,
we obtain the counterterms for the $n$-point $(n=1,...,4)$ vertex
functions in the {\it decoupling limit} as follows:
\bea
\label{eq:counter}
\delta \Gamma_{h^{0}}^{(1)}&=&
\frac{g\mz}{2\cw} \cos 2 \beta
v^2 \left(\sin^2\beta \,\delta Z_{H_{2}}-
\cos^2\beta \,\delta Z_{H_{1}}\right)-v \,\delta M_{12}^2\nonumber\\
&+&\frac{1}{4} \frac{g^2}{\cw^2} v^2 \cos^2 2\beta \,\delta v
- \frac{1}{8} v^3 \cos^2 2 \beta\, \delta G^2\,,\nonumber\\
\delta \Gamma_{h^{0}}^{(2)}&=&\frac{3}{4} \left[
v^2 \cos 2 \beta\,\frac{g^2}{\cw^2}\left(\sin^2\beta \,\delta Z_{H_{2}}-
\cos^2\beta \,\delta Z_{H_{1}}\right)-
\frac{4}{3}\,\delta M_{12}^2 \right.\nonumber\\
&+&\left. \frac{g^2}{\cw^2} \cos^2 2\beta \,v\,\delta v
- \frac{v^2}{2} \cos^2 2 \beta \,
\delta G^2\right]\,,\nonumber\\
\delta \Gamma_{h^{0}}^{(3)}&=& \frac{3}{4} \cos 2 \beta \left[
2v \frac{g^2}{\cw^2}\left(\sin^2\beta \,\delta Z_{H_{2}}-
\cos^2\beta \,\delta Z_{H_{1}}\right)
-\frac{g^2}{\cw^2} \cos 2 \beta \delta v- v\cos 2 \beta \,
\delta G^2\right],\nonumber\\
\delta \Gamma_{h^{0}}^{(4)}&=& \frac{3}{4} \cos 2 \beta \left[
2 \frac{g^2}{\cw^2}\left(\sin^2\beta \,\delta Z_{H_{2}}-
\cos^2\beta \,\delta Z_{H_{1}}\right)-\cos 2 \beta \,\delta G^2\right]\,,
\eea
where we have introduced the abbreviations
\bea
\delta G^2 &\equiv& \delta g^2+\delta g'^2 =
g^2 (2 \,\delta Z_{1}^{W}-3 \,\delta Z_{2}^{W})- 
g'^2 \delta Z_{2}^{B}\,,\nonumber\\
\delta M_{12}^2 &\equiv& \cos^2 \beta \,\delta m_1^2 +
\sin^2 \beta \,\delta m_2^2+\sin 2 \beta \,\delta m_{12}^2\,,\nonumber\\
v \,\delta v &=& v_1 \delta v_1 + v_2 \delta v_2 \,\,\,{\mbox{with}}\,\,\,
v^2=v_1^2+v_2^2\,.
\eea
In the same way, the pseudoscalar-mass counterterm is obtained as
\bea
\label{eq:counmA}
\delta \ma^2 &=& \frac{1}{2} 
\left(
\sin^2 \beta \,\delta m_1^2+\cos^2 \beta \,\delta m_2^2-
\sin 2 \beta \,\delta m_{12}^2 \right)\nonumber\\
&-&\frac{1}{4} \mz^2 \cos^2 2 \beta \left(
\frac{\cw^2}{g^2} \,\delta G^2 +\delta Z_{H_{1}}+\delta Z_{H_{2}}
-2 \frac{\delta v}{v}\right)\,.
\eea

In the on-shell scheme, adopted in this paper,
the counterterms are fixed by imposing the
following renormalization conditions~\cite{Dabelstein,Renor}:\\
-- the on-shell conditions for $M_{W,Z}$ and the electric charge $e$ as
in the minimal $\rsm$,\\ 
-- the on-shell condition for the $A^0$ boson with the pole mass $M_A$,\\
-- the tadpole conditions for vanishing renormalized tadpoles, i.e.\
the sum of the 1-loop tadpole diagrams for $H^0$,
$h^0$, and the corresponding 
tadpole counterterm is equal to zero,\\
-- the renormalization of $\tan\beta $ in such a way 
that the relation $\tan\beta= v_2/v_1$ is valid for the 1-loop 
Higgs minima.

By this set of conditions, the input for the
$\rmssm$ Higgs sector is fixed by the pole mass $M_A$ and $\tan\beta$, together
with the standard gauge-sector input $M_{W,Z}$ and $e$. 

With restriction to the dominant ${\cal O} (m_t^4)$ contributions,
the mass and field counterterms appearing 
in~(\ref{eq:counter})-(\ref{eq:counmA}) have the following structure:
\bea
\label{eq:asympcount}
&& \delta Z_{H_{1,2}} =  0\,,
\,\,\, \delta v = 0\,,\,\,\, 
\delta G^2 =  0\,,\nonumber\\
&& \delta M_{12}^2 = \frac{3}{16 \pi^2}\frac{g^2}{\mw^2}\,
m_t^4\,\left(
\Delta_\epsilon +1-\log\frac{m_t^2}{\mu_{0}^2}\right)\, , 
\nonumber \\
&& \delta \ma^2 = 
  \frac{3}{16 \pi^2}\frac{g^2}{\mw^2}\,
m_t^4\,\cot^2 \beta\,\left(
\Delta_\epsilon +1-\log\frac{m_t^2}{\mu_{0}^2}\right)\, .
\eea 
Now the renormalized vertex functions are obtained as the sum
of the one-loop contributions in~(\ref{eq:oneloop}) and the
counterterms~(\ref{eq:counter}) together with~(\ref{eq:asympcount}).
The renormalized one-point function vanishes, according to 
the corresponding renormalization condition:
$\Delta \Gamma_{h^{0}}^{\,t,\tilde{t}\,(1)}+
\delta \Gamma_{h^{0}}^{\,t,\tilde{t}\,(1)} = 0$. 

The renormalized two-point function is given by
\be
\label{eq:redefinition2p}
\Delta \Gamma_{h^{0}}^{\,t,\tilde{t}\,(2)}+
\delta \Gamma_{h^{0}}^{\,t,\tilde{t}\,(2)} =
-\frac{3}{8 \pi^2}\frac{g^2}{\mw^2}\,m_t^4\,
\log \frac{m_t^2}{m_{\tilde t_1} m_{\tilde t_2}}\, . 
\ee
As expected, the UV-divergence cancels between the one-loop and the
counterterm contributions; however, a logarithmic heavy mass
term, which looks like a non-decoupling effect 
of the heavy particles, remains.
The renormalized two-point function is responsible for a shift in the
pole of the $h^0$ propagator and thus represents the (leading) one-loop
correction to the $h^0$ mass,  
\be
\label{eq:deltamh}
\Delta \Mh^2=-\frac{3}{8 \pi^2}\frac{g^2}{\mw^2}\,m_t^4\,
\log \frac{m_t^2}{m_{\tilde t_1} m_{\tilde t_2}} \, .
\ee
The same expression is obtained from the
results listed in ref.~\cite{Hollikmass}
for the leading one-loop radiative corrections 
from the $t, \tilde{t}-$sector to the light $h^0$ boson
for the  special case of $M_A \gg \mz$ and in the limiting situations
defined in~(\ref{eq:limit}).

For the counterterms to the three- and four-point functions, there
is no ${\cal O} (m_t^4)$ contribution. Hence,
the renormalized $h^0$ three- and four-point vertices,
using the result~(\ref{eq:deltamh}), can be expressed as follows,
\bea
\label{eq:redefinition34p}
&& \Delta \Gamma_{h^{0}}^{\,t,\tilde{t}\,(3)} =
 \frac{3}{v}\,\Delta \Mh^2 
-\frac{3}{8 \pi^2}\frac{g^3}{\mw^3}\,m_t^4 \, , \nonumber \\
&& \Delta \Gamma_{h^{0}}^{\,t,\tilde{t}\,(4)} =
\frac{3}{v^2}\,\Delta \Mh^2 
-\frac{3}{4 \pi^2}\frac{g^4}{\mw^4}\,m_t^4 \, .
\eea
Without the non-logarithmic top-mass term,
the trilinear and quartic $h^0$ self couplings at the
one-loop level have the same form as in~(\ref{treelevelself}),
with the tree-level Higgs mass replaced by the corresponding
one-loop mass
\be
\Mh^2 = \mhtree + \Delta\Mh^2 \, . 
\ee
The terms logarithmic in the heavy-squark 
masses disappear when the vertices are expressed in terms of the
Higgs-boson mass $\Mh$ and, therefore, 
they do not appear directly in related observables, i.e.\
they decouple. 
Moreover, the $h^0$ self-couplings get the form of the self-couplings 
of the SM Higgs boson~(\ref{eq:coupSM}) with $M_H = M_{h^0}$.
The non-logarithmic top-mass  terms are common to both $h^0$ and
$H_{SM}$ (in the SM after renormalization of the trilinear and quartic
couplings).

To make this last point explicit, we give the one-loop ${\cal O} (m_t^4)$ 
contributions for the SM Higgs $n$-point vertex functions, 
which follow from the last four diagrams in Figure~\ref{fig:fdiagrams}
(with $H \equiv H_{\rsm}$ instead of $h^0$ in the external lines)
\bea
\label{SMvert}
&& \Delta \Gamma_H^{(1)} = \frac{3 g}{8 \pi^2 M_W}\, m_t^4 
    \left(\Delta_{\epsilon} - \log \frac{m_t^2}{\mu_0^2} + 1 \right) \, ,
    \nonumber \\
&& \Delta \Gamma_H^{(2)} = \frac{3 g^2}{16 \pi^2 M_W^2}\, m_t^4 
    \left(3\, \Delta_{\epsilon} - 3 \log \frac{m_t^2}{\mu_0^2} + 1 \right) \, ,
    \nonumber \\
&& \Delta \Gamma_H^{(3)} = \frac{3 g^3}{16 \pi^2 M_W^3}\, m_t^4 
    \left(3\, \Delta_{\epsilon} - 3 \log \frac{m_t^2}{\mu_0^2} -2 \right) \, ,
    \nonumber \\
&& \Delta \Gamma_H^{(2)} = \frac{3 g^4}{32 \pi^2 M_W^4}\, m_t^4 
    \left(3\, \Delta_{\epsilon} - 3 \log \frac{m_t^2}{\mu_0^2} -8 \right) \, .
\eea
Differently from the MSSM $h^0$ boson, the 3- and 4-point SM vertices 
are not UV-finite and require renormalization also at the level of the
${\cal O} (m_t^4)$ approximation.
Adding the counterterms, which are derived from the SM Higgs potential
\be
 V = - \frac{\mu^2}{2}\, (v+H)^2 +
 \frac{\lambda}{4} \, (v+H)^4
\ee
via $\rsm$ parameter renormalization 
($\lambda \rightarrow \lambda +\delta\lambda$,
$ \mu^2 \rightarrow \mu^2 +\delta\mu^2$,
$ v \rightarrow v -\delta v$),  
yields the renormalized one-loop vertex functions
\bea
\label{eq:SMvertren}
&& \Delta \hat{\Gamma}_H^{(1)} = 
   \Delta\Gamma_H^{(1)} + \delta\Gamma_H^{(1)}  \, = \,
      \Delta\Gamma_H^{(1)} +  v^3\delta\lambda - v \delta\mu^2 
      -(3v^2\lambda-\mu^2)\, \delta v \, ,
    \nonumber \\
   && \Delta \hat{\Gamma}_H^{(2)} = 
   \Delta\Gamma_H^{(2)} + \delta\Gamma_H^{(2)} \, = \,
      \Delta\Gamma_H^{(2)} +  3v^2\delta\lambda - \delta\mu^2 
    -6v\lambda\, \delta v \, ,
    \nonumber \\
&& \Delta \hat{\Gamma}_H^{(3)} = 
   \Delta\Gamma_H^{(3)} + \delta\Gamma_H^{(3)} \, = \,
      \Delta\Gamma_H^{(3)} +  6 v\, \delta\lambda 
       -6 \lambda\, \delta v \, ,
    \nonumber \\
&& \Delta \hat{\Gamma}_H^{(4)} = 
   \Delta\Gamma_H^{(4)} + \delta\Gamma_H^{(4)} \, = \,
      \Delta\Gamma_H^{(4)} +  6\, \delta\lambda \, .
\eea
The renormalization constant $\delta v$ is determined from the gauge sector
and has no  ${\cal O} (m_t^4)$ contribution, i.e.\ 
$\delta v = 0$. The other renormalization constants $\delta\mu^2$ 
and $\delta\lambda$ have to be determined from the renormalization in
the Higgs sector. The 
corresponding two on-shell conditions are: \\[0.2cm]
-- Tadpole condition: \hspace*{2cm}
    $\Delta \hat{\Gamma}_H^{(1)}\, = \, 0\, ,$ \\[0.2cm]
-- Higgs mass renormalization:  
   $\quad \Delta \hat{\Gamma}_H^{(2)}\, = \, 0\, .$ \\[0.2cm]
Solving these equations yields
\be
\delta \lambda \, =\, \frac{1}{2v^2} \left(\Delta\Gamma_H^{(2)} 
   -\frac{1}{v}\Delta\Gamma_H^{(1)} \right) \, = \,
  \frac{3g^4}{64\pi^2 M_W^4}\, m_t^4
  \left( \Delta_{\epsilon} - \log\frac{m_t^2}{\mu_0} \right) \, ,
\ee
with the expressions in~(\ref{SMvert}) and with $v= 2M_W/g$.
Finally, according to~(\ref{eq:SMvertren}),
one finds  for the renormalized 3- and 4-point vertices
\be
 \Delta \hat{\Gamma}_H^{(3)} = - \frac{3g^3}{8\pi^2 M_W^3}\, m_t^4 \, ,
 \quad \quad 
 \Delta \hat{\Gamma}_H^{(4)} = - \frac{3g^4}{4\pi^2 M_W^4}\, m_t^4 \, ,
\ee
which correspond precisely to the two non-logarithmic terms 
in~(\ref{eq:redefinition34p}).

To summarize this section, 
we conclude that all the ${\cal O} (m_t^4)$
one-loop $\rmssm$ contributions to the $h^{0}$
Green functions in the asymptotic limit either
represent a shift in the $h^{0}$ mass and 
in the $h^0$ triple and quartic self-couplings, which can be
absorbed in $\Mh$, or reproduce the SM top-loop corrections. 
The triple and quartic $h^0$ couplings thereby 
acquire the structure of the SM Higgs-boson self-couplings.
Heavy top squarks 
thus decouple from the low energy theory when the self-couplings are
expressed in terms of the Higgs-boson mass.

\section{Trilinear and quartic $h^{0}$ self-couplings}
\label{sec:tri}

In the previous section, the
results for the one-loop contributions to the
three- and four-point functions were discussed 
considering in the Higgs sector  the {\it decoupling limit} and 
in the squark $\rmssm$ sector
the limit of heavy $\tilde{t}$ masses compared to the electroweak scale,
such that $\tilde{t}_1$ and $\tilde{t}_2$ have masses very close to each
other (cf.~(\ref{eq:limit})). In this section we study
the more general case assuming only that the
stop masses are very heavy compared to the electroweak scale 
(see eq.~(\ref{eq:limitgeneral})), but without further
assumptions on the relative size of the top-squark masses. 
Moreover, also the requirement of 
the {\it decoupling limit} in the Higgs sector is released.
A numerical discussion for the trilinear coupling
shows how fast and to which accuracy
the asymptotic results are achieved, for the cases
specified in~(\ref{eq:limit}) and~(\ref{eq:limit2}).
The numerical analysis of the trilinear self-coupling is appropriately
extended also to the quartic $h^{0}$ self-coupling.

Following the decomposition~(\ref{eq:notG}), 
we write the trilinear self-coupling of the $h^0$ boson
as a sum of the tree-level coupling and the
one-loop radiative correction, 
\be
\label{eq:lambdahhh}
\lambda_{hhh} = \lambda_{hhh}^0 + \Delta \lambda_{hhh}= 
\lambda_{hhh}^0 \left(1+ \frac{\Delta \lambda_{hhh}}{\lambda_{hhh}^0}
\right)\,,
\ee  
where $h\equiv h^0$ and $\lambda_{hhh}^0$ is defined in~(\ref{eq:couptree});
$\Delta \lambda_{hhh}$ is the renormalized one-loop three-point vertex,
\be
\Delta \lambda_{hhh} = \Delta \Gamma_{h^{0}}^{t,\tilde{t}\,(3)}
 \, + \, \delta \Gamma_{h^{0}}^{\,t,\tilde{t}\,(3)} \, , 
\ee
and, accordingly, in similar notation for the quartic coupling.

Concerning the analytic expression for
the one-loop ${\cal O}(m_t^4)$ correction
$ \Delta \lambda_{hhh}$, from the $t,\tilde{t}$-sector, 
we find the result already given in~\cite{RadCorCouplings}
(for $M_{\tilde Q}=M_{\tilde U}$ also in~\cite{osland}),
which can be written in a compact form,
\begin{eqnarray}
\label{eq:completehhh}
\Delta \lambda_{hhh} & = & \frac{3 g^3}{32 \pi^2} 
 \frac{1}{\mw^3}\,m_t^4\, \frac{\cos^3 \alpha}{\sin^3 \beta}
\nonumber \\
& \times & \left\{ 3\log\frac{m_{\tilde t_1}^2 m_{\tilde t_2}^2}{m_t^4}
+ 3 (m_{\tilde t_1}^2 - m_{\tilde t_2}^2)\,C_t F_t
   \log\frac{m_{\tilde t_1}^2}{m_{\tilde t_2}^2} \right. \nonumber\\ 
& & + \left. 2\left(\frac{m_t^2}{m_{\tilde t_1}^2}
\left[1+(m_{\tilde t_1}^2 - m_{\tilde t_2}^2)\,C_t F_t\right]^3
+\frac{m_t^2}{m_{\tilde t_2}^2}
\left[1-(m_{\tilde t_1}^2 - m_{\tilde t_2}^2)\,C_t F_t\right]^3 
-2 \right) \right.\nonumber\\
&+&\left. 3\left(
\frac{M_{\tilde Q}^2-M_{\tilde U}^2}{\msto^2-\mstt^2}\right)^2 
\left[(m_{\tilde t_1}^2 - m_{\tilde t_2}^2)F_t^2
\log\frac{m_{\tilde t_1}^2}{m_{\tilde t_2}^2}+
(m_{\tilde t_1}^2 - m_{\tilde t_2}^2)^2 C_t F_t^3 g_t
\right]\right\},
\end{eqnarray}
with
\begin{eqnarray}
C_t & = & X_t\, /\,(m_{\tilde t_1}^2 - m_{\tilde t_2}^2)\,, \,\,
{\mbox{ with }}\, X_t {\mbox{ defined in }} (\ref{eq:MLRtb})\,,\nonumber\\
F_t & = & (A_t + \mu\tan\alpha)/(m_{\tilde t_1}^2 - m_{\tilde t_2}^2)\,,
\nonumber\\
g_t &=&  2 - 
\frac{m_{\tilde t_1}^2 + m_{\tilde t_2}^2}{m_{\tilde t_1}^2 - 
m_{\tilde t_2}^2} \log\frac{m_{\tilde t_1}^2}{m_{\tilde t_2}^2}\,. 
\label{eq:CFg}
\end{eqnarray}
Notice that the non-logarithmic finite contributions 
to the three-point function from the top-triangle diagram in  
Fig.~\ref{fig:fdiagrams} is also included in~(\ref{eq:completehhh})
(the term with $-2$ in the third line of~(\ref{eq:completehhh})).
It is, however, not taken into account in the figures since it
converges always to the $\rsm$ term.

\begin{figure}[t]
\begin{center}
\begin{tabular}{cc}
\resizebox{8.45cm}{!}{\includegraphics{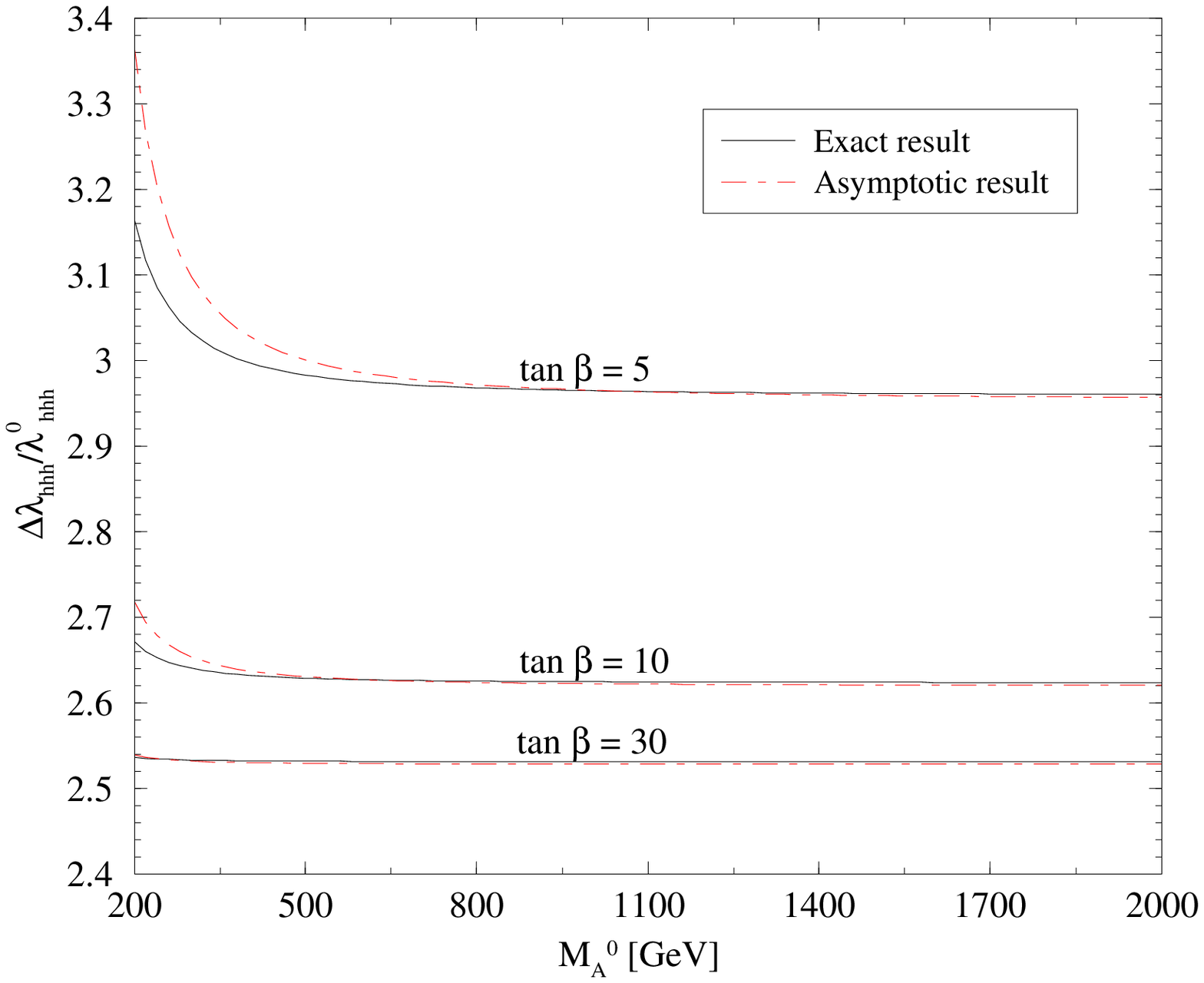}}&
\resizebox{7.7cm}{!}{\includegraphics{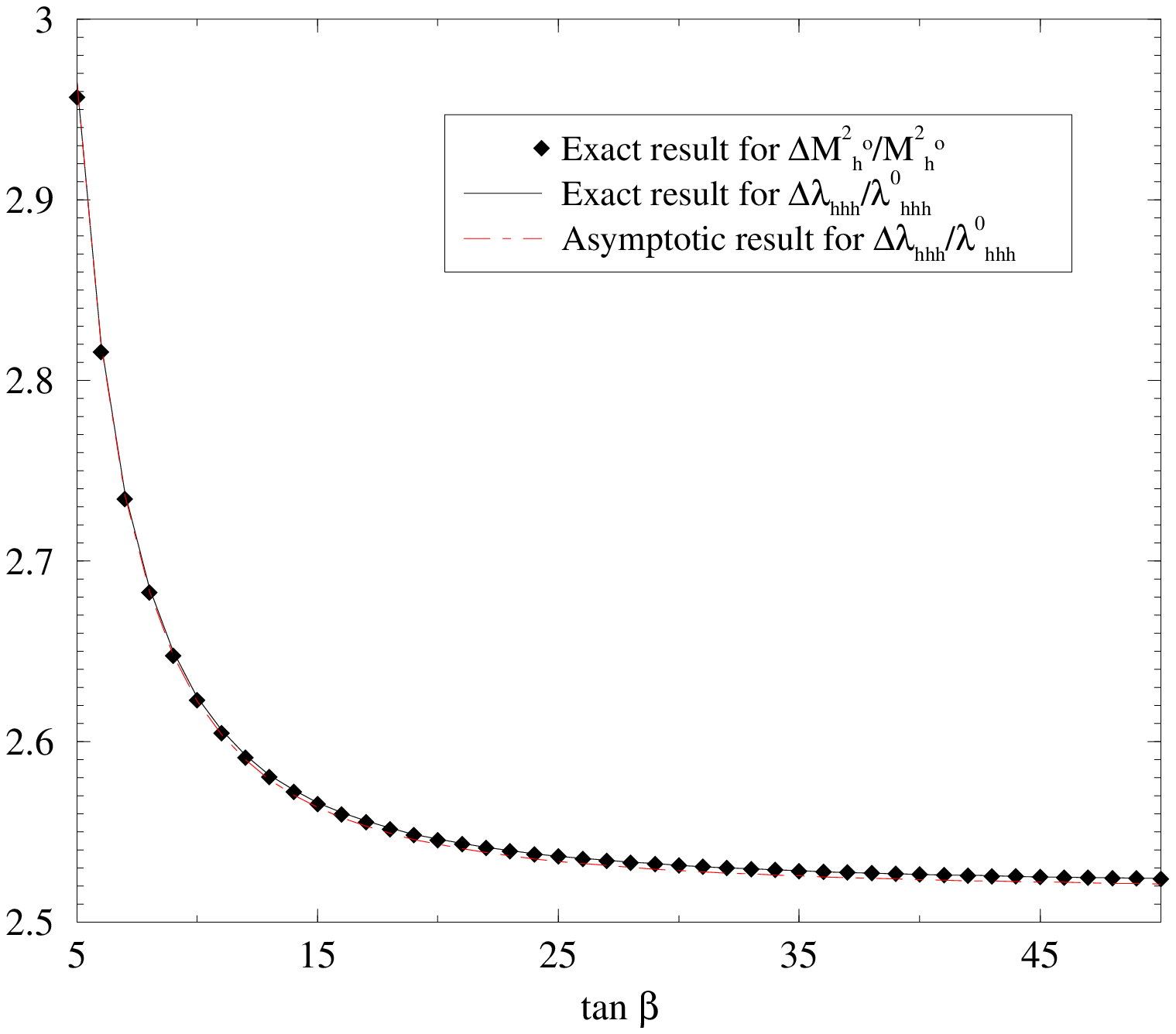}}\\
(a)&(b)
\end{tabular}
\end{center}\vspace*{-0.5cm}
\caption{Exact and asymptotic result of ${\cal O}(m_t^4)$
for \textbf{a)} ${\Delta \lambda_{hhh}}/{\lambda_{hhh}^0}$ 
$\scriptstyle{(h\equiv h^{0})}$ as a function of $\ma$ and
\textbf{b)} ${\Delta \lambda_{hhh}}/{\lambda_{hhh}^0}$ and 
${\Delta \Mh^2}/{\Mh^2}$ as a function of $\tb$, for $\ma=1$ TeV. 
The SUSY parameters have been chosen as in~(\ref{eq:equal}).}
\label{fig:onescale}
\end{figure}

By considering the decoupling limit, which implies 
$\cos\alpha \rightarrow \sin\beta$, 
$\tan\alpha \rightarrow - \cot\beta$, 
and by doing the appropriate expansion in~(\ref{eq:completehhh}) 
for the assumptions given in~(\ref{eq:limitgeneral}),~(\ref{eq:limit}),
one recovers the asymptotic expression 
for the three-point function given in~(\ref{eq:oneloop}).
In order to illustrate also quantitatively how 
the results given in~(\ref{eq:oneloop}) 
and~(\ref{eq:completehhh}) are approached  in the asymptotic limit
of $\Delta \lambda_{hhh}$,
we plot in Fig.~\ref{fig:onescale} the ratio
$\Delta \lambda_{hhh}/\lambda_{hhh}^0$ as function of $\ma$ and $\tb$,
choosing values of the parameters which obey strictly the asymptotic
conditions~(\ref{eq:limit}) for the squark sector:
\be
\label{eq:equal}
M_{\tilde Q}\sim M_{\tilde U}\sim 15 {\mbox{ TeV}}\,,\,\,\,\,\,\,
\mu \sim |A_t| \sim 1.5 {\mbox{ TeV}}\,.
\ee
For definiteness, we also list the following values used for the SM
parameters along all figures in this paper: $G_F=1.16639\times10^{-5}$,
$m_t=175\GeV$, $m_b=4.62\GeV$, $\mz=91.188\GeV$, 
$\mw=80.41\GeV$~\cite{PDG}.

In Fig.~\ref{fig:onescale}\textbf{a}
the variation of the trilinear coupling with $\ma$ is shown 
for different values of $\tb$. Clearly, 
the asymptotic and exact results are in agreement for large 
$\ma$ values, above 500 GeV, depending in detail on $\tb$.
An explicit numerical evaluation of $\Delta \lambda_{hhh}/\lambda_{hhh}^0$
as a function of $\tb$ is presented in Fig.~\ref{fig:onescale}\textbf{b}.
The $A$-boson mass $\ma = 1$ TeV corresponds already to the 
decoupling limit of the Higgs sector, and the various 
results for the triple coupling coincide.
In order to illustrate how well the radiative corrections 
to $\Delta \lambda_{hhh}$ can be described in terms of the corresponding
shift in $\Mh$, asymptotically given in~(\ref{eq:redefinition34p}),
we also display the variation of $\Delta \Mh^2/{\mhtree}$ in this figure.
$\Delta \Mh^2/{\mhtree}$ is represented by
black diamonds; it has been obtained according to the
${\cal O}(m_t^4)$ one-loop Higgs-boson mass results presented 
in~\cite{Dabelstein}. 
The agreement with the vertex corrections is clearly visible.
Therefore, the radiative corrections to $\lambda_{hhh}$, although
large, disappear when $\lambda_{hhh}$ is expressed in terms of $\Mh$. 

So far we have concentrated
on the trilinear $h^{0}$ self-coupling, and we did not give
explicite results for the quartic Higgs boson self-coupling. 
The analytic expressions are quite lengthy and hence we do not list them here.
Numerically, the higher-order contribution to the quartic coupling,
$\Delta \lambda_{hhhh}$, normalized to the tree-level value
$\lambda_{hhhh}^0$ in~(\ref{treelevelself}), show the same behaviour
as the triple coupling in Fig.~\ref{fig:onescale}          
(since the differences are marginal, we do not include an extra figure). 
This is also a numerical proof that the
${\cal O}(m_t^4)$ corrections to the quartic $h^{0}$ self-coupling are
absorbed in the $h^{0}$ mass in the asymptotic limit.

\begin{figure}[t]
\begin{center}
\begin{tabular}{cc}
\resizebox{8.26cm}{!}{\includegraphics{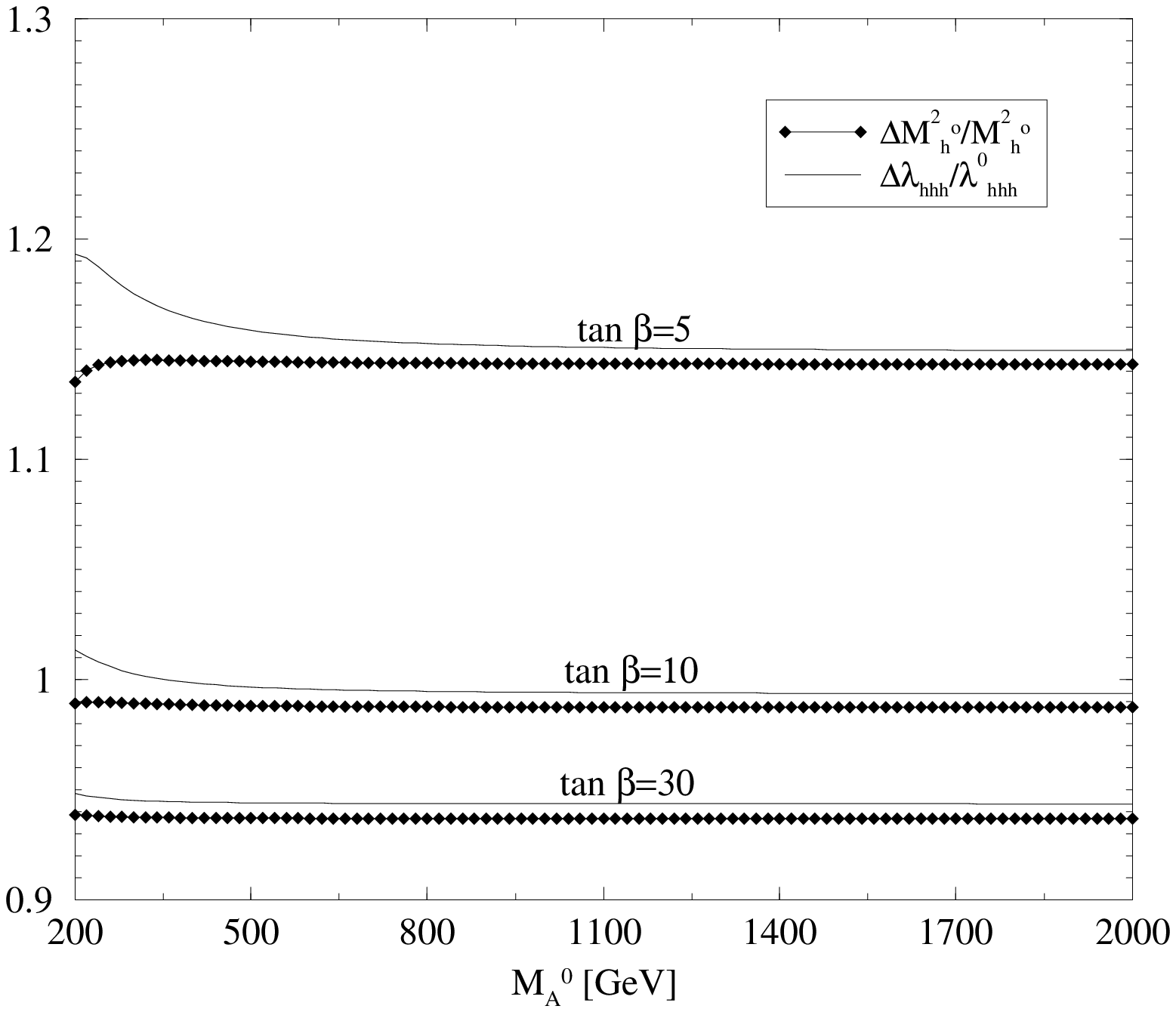}}&
\resizebox{8cm}{!}{\includegraphics{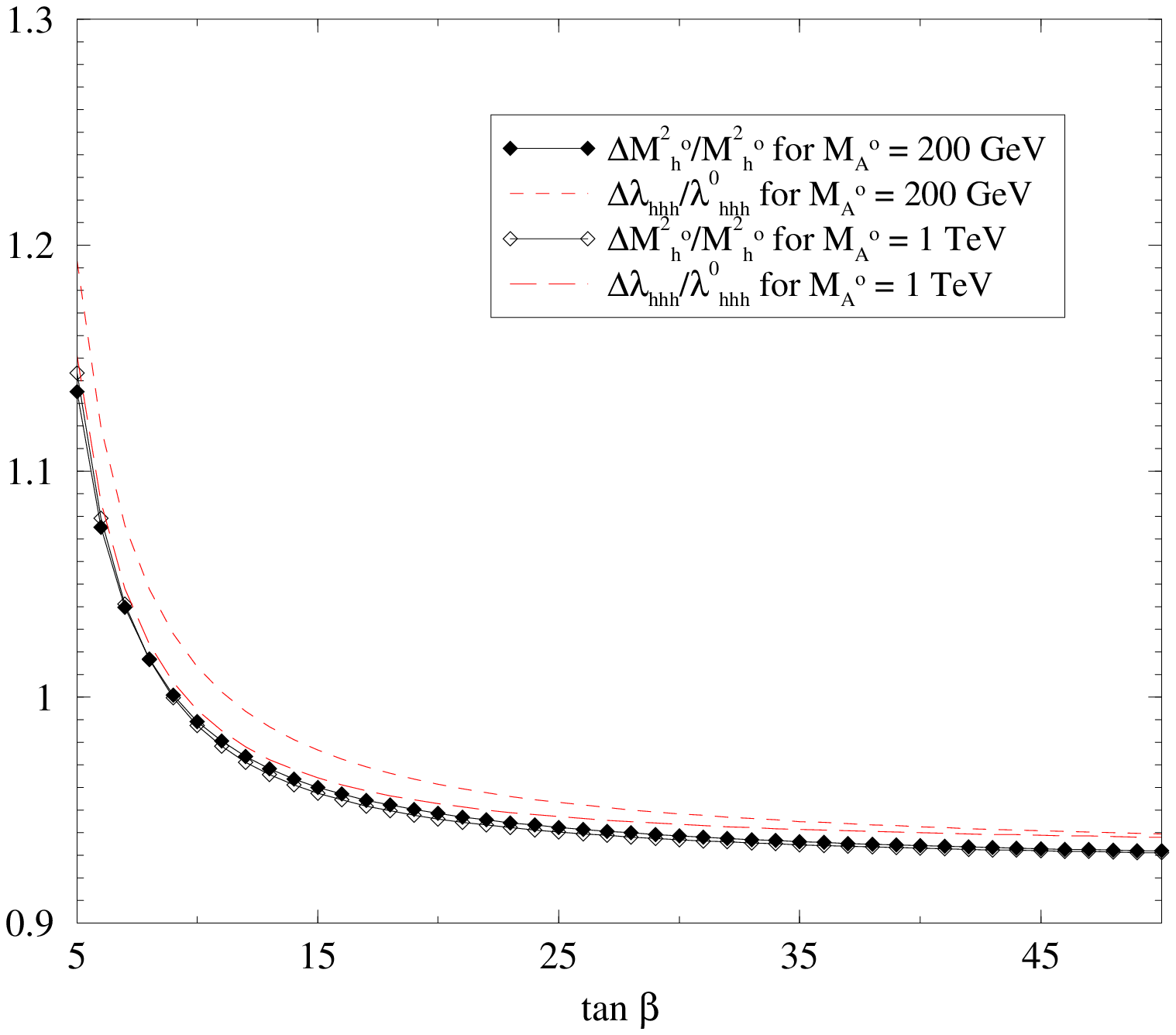}}\\
(a)&(b)
\end{tabular}
\end{center}\vspace*{-0.4cm}
\caption{Leading Yukawa radiative corrections ${\cal O}(m_t^4)$
to the trilinear $h^{0}$ self-coupling and to the $h^{0}$ mass as a
  function of \textbf{a)} \ma, and \textbf{b)} \tb, for choices
  of the SUSY parameters as in~(\ref{eq:difscale}).}
\label{fig:diffscale}
\end{figure}

For the rest of the analysis, we will
consider the limiting situa\-tion in the squark sector
that was specified in~(\ref{eq:limit2}).
In Fig.~\ref{fig:diffscale} we present numerical results for the 
variation of the trilinear coupling, given by the 
expression~(\ref{eq:completehhh}), and for the ${\cal O}(m_t^4)$
$h^{0}$ mass correction, as given in~\cite{Dabelstein}, with $\ma$ and $\tb$. 
The radiative correction to the angle $\alpha$~\cite{Dabelstein} 
is also taken into account. The SUSY parameters have been taken to be
\be
\label{eq:difscale}
M_{\tilde Q}\sim 1 {\mbox{ TeV}}\,,\,\,\,\,\,\,
M_{\tilde U} \sim \mu \sim |A_t| \sim 500 {\mbox{ GeV}}\,.
\ee
With this choice of the $\rsusy$ parameters, the
top-squark masses, $\msto$ and $\mstt$,  are heavy as compared to the 
to the electroweak scale,  but their difference is 
of ${\cal O} (M_{\tilde U})$.

Fig.~\ref{fig:diffscale}\textbf{a} contains the variation of the
trilinear coupling with $\ma$, for different values of $\tb$.
We also give in Fig.~\ref{fig:diffscale} the ${\cal O}(m_t^4)$ 
corrections to
$\Delta \Mh^2/{\mhtree}$ in order to point out how far
the large radiative corrections to the
$h^0$ self-coupling can be absorbed in the $h^{0}$ mass correction, 
$\Delta \Mh^2$.
The relation $\Delta \lambda_{hhh}/{\lambda_{hhh}^0}
\approx \Delta \Mh^2/{\mhtree}$ is only fulfilled up to a small
difference which remains also for large $M_A$. But even in
the most unfavorable cases, namely low $\tb$ and $\ma$ values,
the difference between the $h^0$ mass and
self-coupling at one-loop does not exceed $6\%$
(for $\tb=5$ and $\ma=200$ GeV, it is about $\sim 5\%$). 
The difference decreases for larger values of $\tb$; e.g.\
for $\tb=10$ and $\ma=200$ GeV, the mass and
self-coupling corrections are equal within $\sim 2\%$. This is more 
explicitely displayed in Fig.~\ref{fig:diffscale}\textbf{b}, containing
the variation of $\Delta \lambda_{hhh}/{\lambda_{hhh}^0}$ and 
$\Delta \Mh^2/{\mhtree}$ with $\tb$ for $\ma=200$~GeV and $\ma=1$~TeV.

Therefore, from the numerical analysis one 
can conclude that also for the case of a heavy stop system with large
mass splitting, of the order as the typical SUSY scale, 
the ${\cal O}(m_t^4)$  corrections to
the trilinear $h^{0}$ self-couplings are 
absorbed to a large extent in the loop-induced shift 
of the $h^{0}$ mass, leaving a small difference of only
a few per cent, which can be interpreted 
as the genuine one-loop corrections
when $\lambda_{hhh}$ is expressed in terms of $\Mh$.
Similar results have been obtained also for the 
quartic $h^{0}$ self-coupling, which again 
are close to the ones displayed in Fig.~\ref{fig:diffscale} and hence are
not given in an extra figure.

\section{Conclusions}
\label{sec:conclu}

The ${\cal O}(m_t^4)$  corrections from the $t,\tilde{t}$-sector
to the self-couplings of the 
light CP-even Higgs-boson in the $\rmssm$ have been evaluated. 
We showed analytically that, 
in the limit of large $\ma$ and heavy top squarks,
with $\msto$ and $\mstt$ close to each other, all 
the apparent non-decoupling one-loop effects,
which constitute large corrections to the $h^{0}$ self-couplings, are
absorbed in the Higgs-boson mass $M_{h^{0}}$, 
and the $h^0$ self-couplings get the same form
as the couplings of the SM Higgs boson. 
Therefore, such a heavy top-squark system 
decouples from the low energy theory, 
at the electroweak scale, and leaves behind the SM Higgs sector
also in the Higgs self-interactions. 
 
Other limiting situations where the $\tilde{t}$-mass difference is of the 
order of the $\rsusy$ mass scale have also been investigated.
Similarly to the previous limit, the radiative corrections
to the $h^0$ self-couplings  are large, but their main part can
again be absorbed in the mass $\Mh$. The genuine loop 
corrections to the triple and quartic couplings, after
re-expressing them in terms of $\Mh$, is of the order of a few per cent.
They are largest for low $\tb$ and $\ma$, with typically 5\%.
For large $\ma$, i.e.\ in the decoupling limit of the MSSM
Higgs sector, they decrease to the level of 1\%.
The $h^0$ self-interactions are thus very close to those of 
the SM Higgs boson and would need high-precision experiments
for their experimental verification. 

\vspace{0.4cm}
\section*{Acknowledgments}
 
\noindent The work of S.P. has been
supported by the \textit{Fundaci{\'o}n Ram{\'o}n Areces}.
We thank A.~Dobado, J.~Guasch and 
M.J.~Herrero for valuable discussions and support.
The counterterms have been checked using an independent
Computer Algebra program provided by J.A.~Coarasa.
Support by the European Union under HPRN-CT-2000-00149
is greatfully acknowledged.
 
\begingroup\raggedright

\end{document}